\documentclass[preprint,aps]{revtex4}
\usepackage[dvips]{graphicx}
\usepackage{epsfig}
\usepackage{amsmath}
\usepackage{amsfonts}
\usepackage{amssymb}
\usepackage{dcolumn}
\usepackage{bm}

\begin{document}


\title{Magnetotransport in nanostructures: the role of inhomogeneous currents}

\author{Tiago S. Machado$^{1}$}
\author{M. Argollo de Menezes$^{2}$}
\author{Tatiana G. Rappoport$^{3}$}
\author{Luiz C. Sampaio$^{1}$}%


\affiliation{$^{1}$Centro Brasileiro de Pesquisas F\'{\i}sicas,
  Xavier Sigaud, 150, Rio de Janeiro, RJ, 22.290-180, Brazil
}%

\affiliation{$^{2}$Instituto de F\'{\i}sica, Universidade Federal
  Fluminense, Rio de Janeiro, RJ, 24.210-346, Brazil
}%

\affiliation{$^{3}$Instituto de F\'{\i}sica, Universidade Federal do
  Rio de Janeiro, Rio de Janeiro, RJ, 68.528-970, Brazil
}%


\date{\today}

\begin{abstract}

In the study of electronic transport in nanostructures, electric
current is commonly considered homogeneous along the sample. We use a
method to calculate the magnetoresistance of magnetic nanostructures
where current density may vary in space.  The current distribution is
calculated numerically by combining micromagnetic simulations with an
associated resistor network and by solving the latter with a
relaxation method. As an example, we consider a Permalloy disk
exhibiting a vortex-like magnetization profile.  We find that the
current density is inhomogeneous along the disk, and that during the
core magnetization reversal it is concentrated towards the center of
the vortex and is repelled by the antivortex. We then consider the
effects of the inhomogeneous current density on spin-torque
transfer. The numerical value of the critical current density
necessary to produce vortex core reversal is smaller than the one that
do not take the inhomogeneity into account.

\end{abstract}

\maketitle


\section{Introduction}

Electric transport in magnetic nanostructures is a useful tool both
for probing and for manipulating the magnetization. In the low current
density regime, magnetoresistance curves are useful for probing the
sample's magnetization state while, in the high current density regime,
magnetization patterns can be modified by a spin-transfer torque
\cite{spintorque,spintorque2,yamada}.  Magnetoresistance measurements
have the advantage of being relatively simple and fast, serving as an
efficient magnetic reading mechanism ~\cite{resist,fastread}.

Depending on thickness and diameter, small ferromagnetic disks exhibit
stable topological defects called magnetic vortices
\cite{shinjo,cownburn1}.  These vortices can be manipulated by
picosecond pulses of few (tens of) Oersted in-plane magnetic fields
that switch their polarity \cite{weigand, vansteenkiste, waeyenberger,
hertel, kim, tiago}, making them good candidates for elementary data
storage units \cite{vansteenkiste}.  

For their use as storage units, the most viable form of manipulation
of the magnetization is through spin-torque transfer, with the
injection of high density electrical currents~\cite{spintorque}. The
effect of these currents in the magnetization dynamics is described
theoretically by the incorporation of adiabatic and non-adiabatic
spin-torque terms in the Landau-Lifshitz-Gilbert (LLG)
equation~\cite{zhang,tiaville}. These two terms are proportional to
the injected current density and it is normally considered an
homogeneous current distribution inside the disk. Although theoretical
predictions using this approach agree qualitatively with experimental
results, there is a lack of quantitative agreement between theoretical
and experimental results regarding to the current densities necessary
to modify the magnetic structures~\cite{beach,yamada08,yamada10}.

In this paper we investigate the effect of non-uniform current
distributions on electronic transport and spin-torque transfer in
ferromagnetic systems exhibiting vortices.  We calculate numerically
the magnetoresistance (MR) and local current distribution of a
ferromagnetic disk by separating the timescales for magnetic ordering
and electronic transport. We consider an effective anisotropic
magnetoresistance (AMR) that depends on the local magnetization. We
discretize the disk in cells and solve the Landau-Lifshitz-Gilbert
(LLG) equation ~\cite{LLG} numerically with fourth-order Runge-Kutta
\cite{nrecipes}, obtaining the magnetization profile of the disk. This
pattern is used to calculate the magnetoresistance of each cell as a
fixed current $I$ is applied at two symmetrically distributed
electrical contacts, resulting in a voltage drop and an inhomogeneous
current distribution along the disk.

This method couples the electric and magnetic properties of the
metallic nanomagnets and can be used to analyze the effect of
inhomogeneous current distributions in different contexts. First, we
discuss the limit of low current density where transport measurements
can be used to probe the magnetic structure. We compare the magnetic
structure with magnetoresistance curves and show how magnetoresistance
measurements could be interpreted to obtain information on the
magnetization profile and its dynamics during the vortex core
magnetization reversal.  Moreover, we discuss the consequences of a
non-homogeneous current distribution on spin-torque transfer and find
that the critical current density that produces vortex core reversal
is reduced by one order of magnitude whenever such non-inhomogeneity
is taken into account. This result can be seen as a new route to
understand why experimental values of the critical current densities
are usually lower than the ones obtained in LLG
calculations~\cite{beach,yamada08,yamada10}.

This article is organized as follows: In section \ref{s2} we discuss
the model and method for the calculation of the magnetoresistance and
current distribution.  In section \ref{s3}, we exemplify the
calculations by considering the magnetoresistance and current
distributions of a Permalloy disk exhibiting a magnetic vortex. In
section \ref{s4}, we study the consequences of a non-homogeneous
current distribution on spin-torque transfer. In section \ref{s5} we
summarize the main results.

\section{Magnetoresistance and current distribution calculations\label{s2}}
 
Let us consider a $36$nm-thick Permalloy disk with a diameter of $300$
nm discretized into a grid of $4\times 4 \times 4$nm$^3$ cells.  The
dynamics of the magnetization vector associated with each cell is
given by the Landau-Lifshitz-Gilbert equation, which we numerically
integrate with fourth-order Runge-Kutta and discretization step
$h=10^{-4}$ ~\cite{nrecipes}.  The parameters associated with the LLG
equation are the saturation magnetization $M_s=8.6 \times 10^5$A/m,
exchange coupling $A=1.3\times10^{-11}$J/m and Gilbert damping
constant $\alpha=0.05$ \cite{tiago}.

\begin{figure}[h]
    \includegraphics[width=0.7\columnwidth]{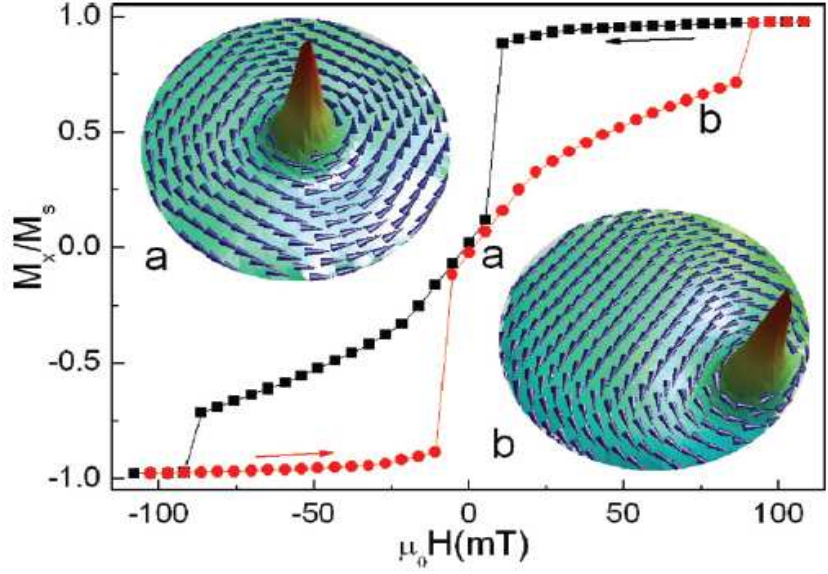}
    \caption{Magnetic hysteresis obtained with micromagnetics
      simulation of a Py disk with a diameter of $300$ nm and
      thickness of $36$ nm subject to a static, in-plane magnetic field
      $H$. Two configurations for the vortex core, corresponding to
      different external fields ($0$ and $75$ mT), are also depicted.}
    \label{figure1}
\end{figure}

By varying the external in-plane magnetic field $H$ from negative to
positive saturation we obtain a hysteresis curve, as depicted in
Fig.~\ref{figure1}, which is consistent with experimental observations
\cite{cownburn1}. As shown in Fig.~\ref{figure1}(a), in static
equilibrium and in the absence of magnetic fields, a vortex structure
with a core magnetized perpendicular to the disk plane is formed in
the center of the disk. If a small in-plane magnetic field $H$ is
applied, the core is displaced from the center
(Fig.~\ref{figure1}(b)). At a critical field $H_{c1}$ the vortex is
expelled from the disk, resulting in a discontinuity in the hysteresis
loop. As the external field $H$ is lowered back, the vortex structure
reappears, but at a lower field $H_{c2}<H_{c1}$.

\begin{figure}[!h]
    \includegraphics[width=0.4\columnwidth]{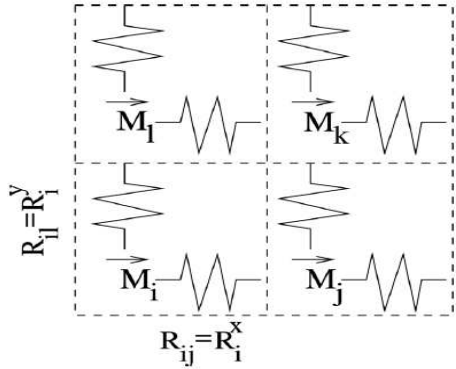}
    \caption{Original cells used in the LLG simulation with
      the associated resistance network.}
    \label{figure2}
\end{figure}

In order to investigate the electronic transport on the nanomagnetic
disk we consider the magnetization profile $\{\vec{M}_{i}\}$, obtained
as the stationary solution of the LLG equation, as a starting point to
calculate the magnetoresistance $R_{i}$ in each cell $i$ of the
disk. It is well established that in relatively clean magnetic metals
the main source of magnetoresistance is the anisotropic
magnetoresistance (AMR)~\cite{AMR}, which can be expressed as $\rho =
\rho^{\perp} + (\rho^{\parallel} - \rho^{\perp}) {\cos^2{\phi}}$,
where $\phi$ is the angle between the local magnetization and the
electric current and $\rho^{\perp}$ and $\rho^{\parallel}$ are the
resistivities when the magnetization is perpendicular and parallel to
the current, respectively. We decompose the current into orthogonal
components $x$ and $y$ such that if the normalized projection of the
magnetization $\vec{M_i}$ on the current direction $\hat{u}$ ($u=x,y$)
is $m^u_{i} = \cos\phi$, and the cell geometrical factor is taken into
account, the magnetoresistance $R_i$ is split into orthogonal
components as $R^u_i = R^{\perp}_i + (R^{\parallel}_i - R^{\perp}_i)
{(m^u_{i})}^2$ in every cell $i$ of the disk (Fig.~\ref{figure2}).
Thus, we obtain a resistor network where the resistances depend on the
local magnetization and are assumed to be approximately constant at
the time scale of electronic scattering processes.

Guided by recent experiments ~\cite{kasai} we allow a constant current
$I$ to flow along the disk by attaching symmetrically placed
electrodes on it (see Fig.~\ref{figure3}). The voltage drop along the
resistors and the associated current map of the disk are obtained by
solving Kirchoff's equation iteratively at each node of the grid with
a relaxation method~\cite{cou,gould}:

\begin{equation}
V_i^{(n+1)}=\left(\sum_{\left<j\right>}1/R_{ij}\right)^{-1}\left(\sum_{\left<j\right>}\frac{V^{(n)}_j}{R_{ij}}+b_i\right),
\label{iteration}
\end{equation}

\noindent where $R_{ij}$ is $R^x_{i} (R^y_{i})$ if $i$ and $j$ are
horizontal (vertical) neighbors and $b_i$ is the boundary current,
assumed to be $I (-I)$ at the leftmost (rightmost) cells and zero
otherwise (see Fig.~\ref{figure2}). $V_i^n$ is the voltage at site $i$
after $n$ iterations and the sums run over the nearest-neighbors
$\left < j \right>$ of node $i$.  Starting with a random initial
condition $ \lbrace {V_i}^{(0)} \rbrace $ at each site we iterate
Eq.\ref{iteration} until each ${V_i}^{(n)}$ becomes stationary (within
$9$ decimal digits precision). After convergence we calculate the
equivalent resistance, the ratio $R_{eq}=\Delta V/I$ between the
voltage drop $\Delta V$ between the electrodes, given by

\begin{equation}
\Delta V=\sum_{i\|b_{i}=I}V_i-\sum_{j\|b_{j}=-I}V_j, 
\end{equation}

\noindent and the current $I$ entering the disk. 

\section{Magneto-structure and  magnetoresistance\label{s3}}

\subsection{Hysteresis and magnetoresistance}

\begin{figure}[h]
    \includegraphics[width=0.5\columnwidth]{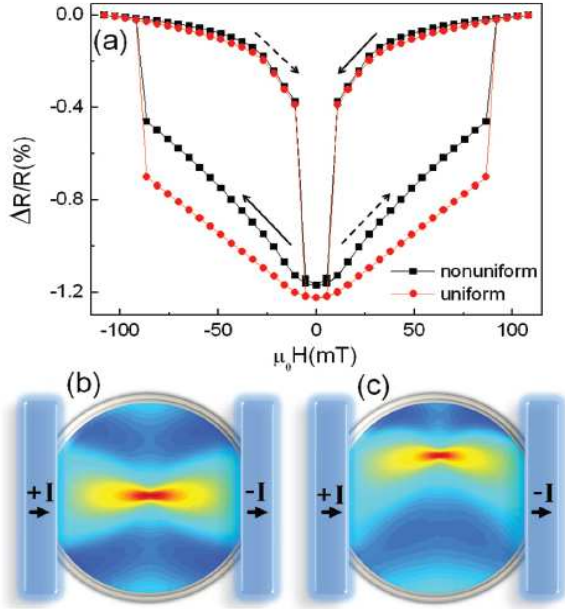}
   \caption{(a) Magnetoresistance for the magnetic configurations
     obtained in Fig~\protect\ref{figure1} for uniform (circle) and
     non-uniform (square) current distribution.  Bottom: electric
     current map for (b) zero field and (c) $H=75$ mT. The red (blue)
     color corresponds a current density which is about $1-2\%$ larger
     (smaller) than the uniform current at the saturation field.}
    \label{figure3}
\end{figure}

In order to obtain the magnetoresistance curves, the calculation
discussed in the previous section is performed at different
fields. Magnetoresistance and current distribution for the same points
of the hysteresis loop in Fig.~\ref{figure1} are depicted in
Fig.~\ref{figure3}.  Fig.~\ref{figure3}(a) displays magnetoresistance
curves for both homogeneous (without using the resistance
network~\cite{renato}) and non-homogeneous current distributions. The
vortex expulsion and its formation at a different critical field are
clearly identified and, with $\rho^{\parallel}=155 \Omega\cdot$nm and
$\rho^{\perp}=150 \Omega\cdot$nm, we obtain a MR of $1.2\%$ for the
non-homogeneous distribution, which is a typical value found in
experiments~\cite{kasai, vavassori}.

One also observes that magnetoresistance curves for uniform and
non-uniform current distributions differ significantly, the latter
being more comparable to experimental results with same contact
geometry~\cite{kasai}. As expected, a homogeneous current
overestimates the magnetoresistance, since the current will flow thru
regions of high resistance, whereas if the current found by solving
Laplace's equation on the associated resistor network, one finds a
preferential path (higher current density) on regions of low
resistance. This difference is more pronounced in the presence of a
vortex, since the magnetization of the disk is highly non-homogeneous
on such configuration.

In the light of the discussion above, one sees in Fig.~\ref{figure3}
(b)-(c) that the current is not homogeneously distributed inside the
disk, being stronger towards the center of the vortex core. In the
center of the disk the magnetization either points in $\hat{z}$
direction, perpendicular to the direction of current flow, or loops
about the vortex core. In both cases, the current has a path where its
direction is always perpendicular to the magnetization, reducing the
local magnetoresistance. Above the saturation field, the magnetization
is uniform and at the disk center the same happens to the current. The
red (blue) region has a current density $1-2\%$ larger (smaller) than
the current $I$ at saturation. This effect might be enhanced if other
sources of magnetoresistance are considered, as giant
magnetoresistance for example. Similar approaches, using different
sources of magnetoresistance and geometries had been used to calculate
the magnetoresistance in nanomagnets ~\cite{vavassori, Li, Bolte,
  Holz, Ohe, bogart}.

\subsection{Dynamics}

\begin{figure}[!h]
  \includegraphics[width=0.5\columnwidth]{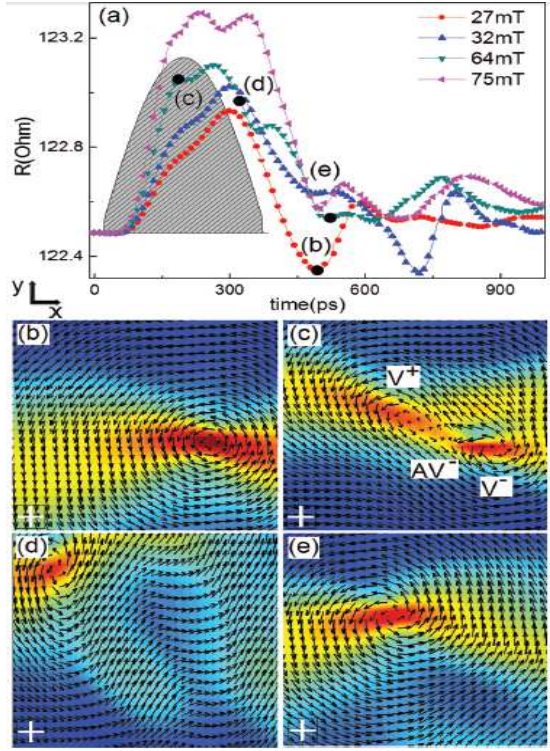}
  \caption{(a) Evolution of magnetoresistance after the application of
    pulsed, in-plane magnetic fields (the shape is shown in grey) with
    different intensities. Snapshots of magnetization (arrows) and
    current distribution (color map) for pulse fields without (b) and
    with (c)-(e) vortex core magnetization reversal.  The white cross
    shows the position of the disk center. Both current and magnetic field are applied in the $\hat{x}$ direction }
  \label{figure4}
\end{figure}

Next, we study the dynamics of the vortex core magnetization reversal
by the application of short in-plane magnetic fields.  Under a pulsed
in-plane magnetic field or spin polarized current excitation, the
vortex with a given polarity (V$^+$) dislocates from the center of the
disk with nucleation of a vortex (V$^-$)-antivortex (AV$^-$) pair with
opposite polarity after the vortex attains a critical velocity of
rotation about the disk center \cite{yamada, gus}. The original V$^+$
then annihilates with the AV$^-$, and a vortex with reversed core
magnetization (V$^-$) \cite{vansteenkiste} remains. If a low-density
electronic current is made to flow through the sample (without
disturbing the magnetization dynamics), we observe changes in the
magnetoresistance, as the vortices nucleate and annihilate.  We depict
in Fig.~\ref{figure4}(a) the dynamics of the magnetoresistance as a
pulsed, in-plane magnetic field is applied in the $\hat{x}$ direction
at $t=20$ ps for different pulse intensities. The pulses have their
shape sketched in grey in Fig.~\ref{figure4}(a) with full width at
half maximum of t=250 ps.  Depending on the pulse intensity, the
vortex core magnetization does not reverse at all ($\mu_0 H<43$ mT),
reverses once ($54$ mT$<\mu_0H<64$ mT) or multiple times ($\mu_0 H>64$
mT)~\cite{kim,tiago}.

During the application of a field pulse in the $\hat{x}$ direction,
i.e., parallel to the current flow, the vortex core is pushed to the
$\hat{y}$ direction breaking the rotation symmetry of the disk's
magnetization, increasing both the total $m_x$ component and the
disk's equivalent resistance (see Fig.~\ref{figure4}(a)).  At $t=340$
ps the field is practically zero and, from the decay of
magnetoresistance to its equilibrium (initial) value, one can infer
whether there was reversal of the vortex core polarization or not: for
pulses that induce reversal, the value of the magnetoresistance just
after the pulse is always larger than its initial value.  If there is
no reversal the magnetoresistance attains a minimum value that is
lower than its initial value, i.e., before the application of the
pulse, and oscillates about it.

In Figs.~\ref{figure4}(b)-(e) we depict snapshots of current (color
map) and magnetization (arrows) distributions at time steps marked
with black dots in Fig.~\ref{figure4}(a), in situations with or
without vortex core magnetization reversal.  Whenever the pulse
decreases its intensity, the total $m_x$ component and the equivalent
resistance of the disk follow the same pattern (although with some
time delay), because the vortex core tends to return to the disk
center, where $m_x = 0$.  Panel (b) shows the current distribution and
magnetization at a moment corresponding to the minimum of the
resistance curve, for a field intensity $\mu_0 H=27 mT$, for which
there is no vortex core reversal. There is a large region with $m_y$
magnetization (and small $m_x$) in the center of the disk. This
region, together with the vortex core, creates a low resistance path
for the electronic current, decreasing the equivalent resistance
towards a value below the equilibrium resistance. Panels (c), (d) and
(e) show magnetization and current distributions at different moments
of the vortex core magnetization reversal for a situation where there
is a single reversal ($\mu_0H=64 mT$).  In panel (c) we depict the
current distribution at the exact moment of nucleation of the
V$^-$AV$^-$ pair, the initial stage of vortex core magnetization
reversal.  Panel (d) shows the spin waves emitted just after the
V$^+$AV$^-$ annihilation, a process that occurs with energy
dissipation. Such energy loss drives the vortex core to the disk
center along with some small oscillations, mainly due to reflections
of spin waves at the edges of the disk. It turns out that the
resistance follow equivalent behavior: it decreases towards the
initial resistance value and remains always above it.  Panel (e) shows
the current distribution after the field pulse has vanished.  As can
be seen, the time dependent resistance curves can give us an
indication of the vortex reversal process.

\begin{figure}[h]
  \includegraphics[width=0.7\columnwidth]{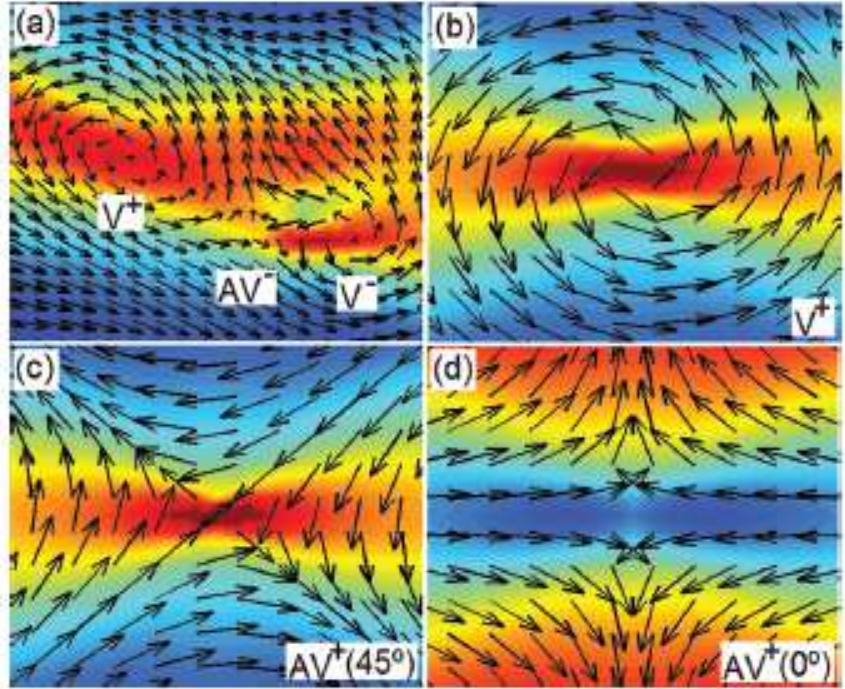}
  \caption{ (a)-(d) Snapshots of magnetization (arrows) and current
    distribution (color map) of (a) the vortex core magnetization
    reversal process at $t= 214$ ps, (b) a vortex and its associated
    current distribution, and (c)-(d) antivortices rotated by 45
    degrees with respect to each other and their associated current
    distribution. Depending on the relative orientation of the
    antivortex it can either focus (c) or repel (d) currents away from
    the center of the core.}
    \label{figure5}
\end{figure}

Let us discuss in further details the interplay between magnetization
pattern and current distribution.  In Fig.~\ref{figure5} (a) we show a
snapshot of the current distribution during the vortex core
magnetization reversal process, with the V$^+$ and the V$^-$-AV$^-$
pair with negative polarity. As shown in Figs. \ref{figure3},
\ref{figure4} and \ref{figure5}(b) the current is pushed to the vortex
core, where $m_x=0$ and, consequently, the local resistance is
minimum.  With the nucleation of the AV$^-$ vortex (Fig.~\ref{figure5}
(a)), $m_x$ gets larger than zero around it, with $m_y \to 0$. As
current flows in the $\hat{x}$ direction, it is {\it repelled} from
the antivortex core.

In the latter analysis we considered a particular orientation of the
AV. However, as can be seen in Figs.~\ref{figure5}(c) and (d),
depending on their orientation, antivortices can either attract (in
the first case) or repel currents (in the latter case). Vortices are
rotation invariant, and alway attract current towards their cores. It
is important to point out that this difference in current
distributions might have important consequences in the high-density
current spin-torque transfer acting on either a vortex or an
antivortex. For instance, although the inversion process through
spin-torque for an AV is equivalent to the one for a V, we should
expect different current densities in each one, since currents can
only penetrate the AV core at a particular orientation.

\section{Spin-torque transfer\label{s4}}

In this section, we discuss the consequences of inhomegenous currents
in the spin-torque transfer.  In order to check how the current
distribution is incorporated in the spin-torque terms of the modified
LLG equation, we need to review a few steps of their derivations. It
is important to note that in our approach, the only source of
non-homogeneous current distribution is the anisotropic
magnetoresistance, as discussed in section \ref{s2}. All other effects
are neglected.

The itinerant electrons spin operator satisfies the continuity equation 
\begin{equation}
  \dfrac{d}{dt}\langle\textbf{s}\rangle+\nabla\cdot\langle\hat{\textbf{J}}\rangle=-\dfrac{i}{\hbar}(\langle[\textbf{s},H]\rangle) 
  \label{eq_d_e1}
\end{equation}
where $\hat{\textbf{J}}$ is the spin current operator. The Hamiltonian
$H$ is the s-d Hamiltonian
($H_{sd}=-J_{ex}\textbf{s}\cdot\textbf{S}$), where $\textbf{s}$ and
$\textbf{S}/S=-\textbf{M}/M_{s}$ are the spins of itinerant and
localized electrons, and $J_{ex}$ is the exchange coupling strength
between them. We define spin current density
$\textbf{J}=\langle\hat{\textbf{J}}\rangle=-(g\mu_{B}P/eM_{s})\textbf{j}_{e}(\textbf{r})\otimes\textbf{M})$,
where $\textbf{j}_{e}(\textbf{r})$ is the current density, and the
electron spin density is given by
$\textbf{m}=\langle\textbf{s}\rangle$\cite{zhang}. We use the same
approximations, previously used to calculate the spin-torque
~\cite{zhang,tiaville}, with the new ingredient of non-homogeneous
current density. We obtain
\begin{equation}
  \begin{array}{ll}
     \dfrac{d}{dt}\textbf{m}=\dfrac{\mu_{b}P}{eM_{s}}[\textbf{M}(\nabla\cdot\textbf{j}_{e}(\textbf{r}))+(\textbf{j}_{e}(\textbf{r})\cdot\nabla)\textbf{M}]\\
\hspace{0.9cm}-\dfrac{J_{ex}S}{M_{s}}\textbf{m}\times\textbf{M},\\
    \end{array}
    \label{eq_fi_e_c}
\end{equation}
where, $\textbf{M}$ is the matrix magnetization, $g$ is the Land\'e
factor splitting, $\mu_{B}$ is the Bohr magneton, $P$ is the spin
current polarization of the ferromagnet, $e$ is the electron
charge. From the continuity equation for charges, the term containing
$\nabla\cdot\textbf{j}_{e}(\textbf{r})$ is always zero, even if the
current density is not constant. As discussed previously, the same
divergent is used to determine the current distribution in section
\ref{s2}. This expression is exactly the same expression obtained
previously but with $\textbf{j}_{e}(\textbf{r})$ in the second term of
the right side of the equation varying with $\textbf{r}$. This current
distribution is introduced at the modified LLG that considers
spin-torque transfer. Therefore, we obtain a spin-torque transfer
where the current distribution is not uniform.

To consider spin-torque transfer effects we include adiabatic and
non-adiabatic spin torque terms in the LLG equation,
\begin{equation}
  \begin{array}{ll}
    \dfrac{d}{dt}\textbf{m}=-\gamma_{0}\textbf{m}\times\textbf{H}_{eff}+\alpha\textbf{m}\times\dfrac{d}{dt}\textbf{m}\\
\hspace{1.2cm}                   -(\textbf{u}\cdot\nabla)\textbf{m}+\beta\textbf{m}\times[(\textbf{u}\cdot\nabla)\textbf{m}],\\
    \end{array}
\label{LLGmod}
\end{equation}
where, \textbf{m}=\textbf{M}/$M_{s}$ is the normalized local
magnetization, $\alpha$ is a phenomenological damping constant,
$\gamma_{0}$ is the gyroscopic ratio, \textbf{H}$_{eff}$ is the
effective field, which is composed of the applied external field, the
demagnetization field, the anisotropy field and the exchange field.
The first term describes the precession of the normalized local
magnetization about the efective field. The second term describes the
relaxation of the normalized local magnetization and $\beta$ is a
dimensionless parameter that describes the strength of the
non-adiabatic term, which we consider to be 0.5~\cite{tiaville,
  Heyne}. The velocity
\textbf{u(\textbf{r})}=$(gP\mu_{B}/2eM_{s})$\textbf{j}$_{e}$(\textbf{r})
is a vector pointing parallel to the direction of electron flow and
\textbf{j}$_{e}$(\textbf{r}) is calculated using the procedure
discussed in section \ref{s2}.

\begin{figure}[!h]
  \includegraphics[width=0.7\columnwidth]{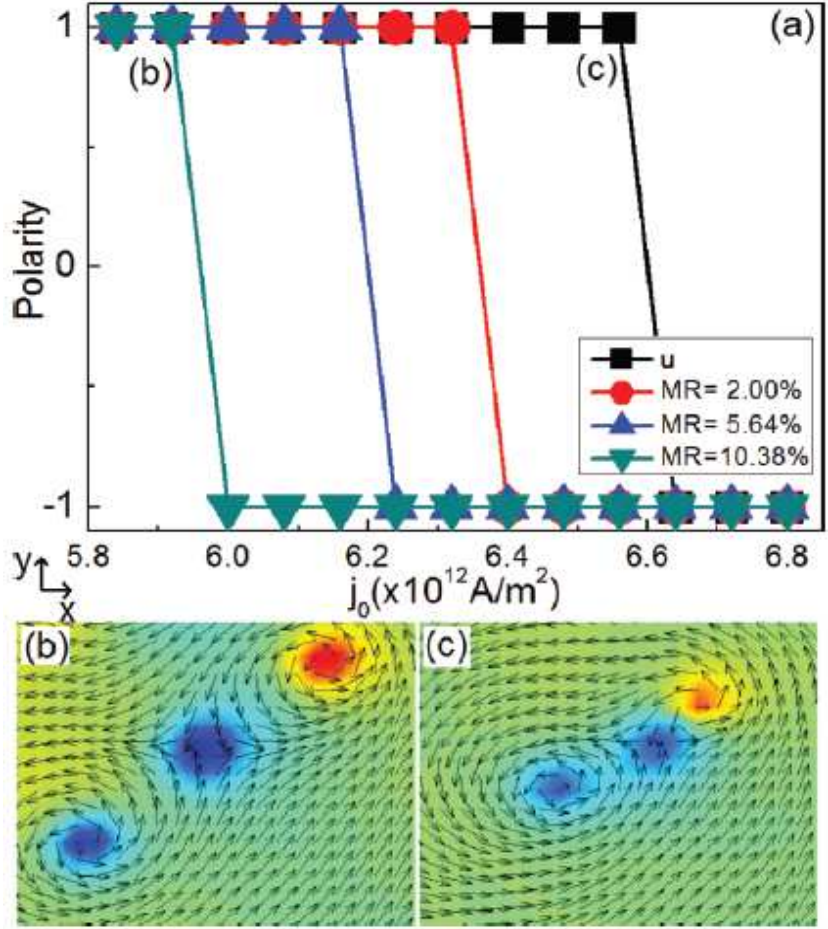}
  \caption{(a): Core polarity as a function of current density for
    homogeneous (squares) and non-homogeneous current distributions
    (with different MR). (b) and (c): magnetization profiles during
    the inversion process at the critical current for both
    non-homogeneous (b) and homogeneous (c) current distributions. The
    color map represent the out-of-plane magnetization, $m_z$, and the
    arrows represent the in-plane component.}
  \label{core_polarity}
\end{figure}

To explain the importance of our assumption about the current
distribution let us analyze the critical current density
$\textbf{j}_{e}^{c}$, the minimal current density needed to produce a
vortex core reversal. For this purpose, we simulated the magnetization
dynamics of a system subjected to a DC current with the modified LLG
equation (equation ~\ref{LLGmod}). In Fig.~\ref{core_polarity}(a) one
sees the vortex core polarity as a function of current density
$\textbf{j}_{e}$. The different curves represent situations of
homogeneous current (squares) and of three different values of AMR
where the magnetoresistance ranges from $2\%$ to $10\%$. Such AMR, as
discussed in the previous sections, shape the degree of current
inhomogeneity throughout the disk. One can see that the critical
current density $\textbf{j}_{e}^{c}$ in our model is $3-10\%$ smaller
than the one obtained for uniform currents.  These results suggest a
new route, together with the non-adiabatic term, to explain the discrepancy 
between experimental results and
theoretical calculations of the critical current density
$\textbf{j}_{e}^{c}$.

Our analysis might also have important technological implications,
since we observe a (almost linear) correlation between the current
density necessary to produce a core inversion and the anisotropic
magnetoresistance of the material. Thus, by increasing the AMR of the
sample, one can decrease the critical current $\textbf{j}_{e}^{c}$,
which is strongly desirable in memory devices for the sake of low
energy comsumption and minimal heat waste.

Panels (b) and (c) of Fig.~\ref{core_polarity} show the magnetic
configurations for the moment just before the V$^{+}$-AV$^{-}$
annihilation, at the critical current density, for the model with
non-homogeneous and homogeneous current distributions,
respectively. In the case of inhomegeneous currents, the fact that a
V(AV) attracts (\textit{repels}) the current affects the velocity and
separation distance of the V-AV pair during the vortex core
reversal. As antivortices \textit{repel} currents, the current density
at their core is smaller, making them slower than vortices. As a
result, after the nucleation of the V$^{-}$-AV$^{-}$ pair their
separation occurs faster than in the case where the current density in
the center of a V or an AV is the same, as usually considered in
micromagnetic simulations.

\vskip 1 pt

\section{Conclusions\label{s5}}

We performed a realistic calculation of magnetoresistance effects in
magnetic nanostructures subject that takes into account 
inhomogeneous current densities. For that purpose, we adapted a numerical relaxation scheme for the Laplace equation to the solution of the LLG equation for the magnetization
profile along a Permalloy disk. Our results suggest that resistance
measurements might be useful to probe the dynamics of vortex core
magnetization reversal, induced by short in-plane magnetic pulses. 
Moreover, we note that the difference between current distributions close to
vortices and anti-vortices have significant consequences for the
spin-torque transfer effect. The inhomogeneous current
distribution inside the magnet reduces substantially the critical current
density necessary to produce a vortex core reversal. We
conclude that materials with large anisotropic magnetoresistance need
lower current densities to modify their magnetic structure, a much
deasireable feature for most modern memory devices.

\section*{Acknowledgements}
This work was supported by CNPq and FAPERJ. LCS and TGR
acknowledge the ``INCT de Fot\^onica'' and ``INCT de 
Informa\c c\~{a}o Qu\^antica'', respectively, for financial support.

\end{document}